\documentclass[12pt]{article}

\input epsf

\begin{document}

\begin{titlepage}
\begin{flushright}
CALT-68-2518\\
\end{flushright}

\begin{center}
{\Large\bf $ $ \\ $ $ \\
Notes on fast moving strings}\\
\bigskip\bigskip\bigskip
{\large Andrei Mikhailov\footnote{e-mail: andrei@theory.caltech.edu}}
\\
\bigskip\bigskip
{\it California Institute of Technology 452-48,
Pasadena CA 91125 \\
\bigskip
and\\
\bigskip
Institute for Theoretical and 
Experimental Physics, \\
117259, Bol. Cheremushkinskaya, 25, 
Moscow, Russia}\\

\vskip 1cm
\end{center}

\begin{abstract}
We review the recent work on the mechanics of fast moving
strings in anti-de~Sitter space times a sphere and discuss the role
of conserved charges. An interesting relation between the local
conserved charges of rigid solutions was found in the earlier
work. We propose a generalization of this relation for arbitrary solutions,
not necessarily rigid. We conjecture that an infinite
combination of local conserved charges is an action variable generating
periodic trajectories in the classical string phase space. 
It corresponds to the length
of the operator on the field theory side.
\end{abstract}

\end{titlepage}

\section{Introduction.}
The AdS/CFT correspondence
is a strong-weak coupling duality. 
Weakly coupled Yang-Mills
is mapped to the string theory on the highly curved 
AdS space. When AdS space is highly curved,
the string worldsheet theory becomes strongly coupled. 
Therefore, the weakly coupled Yang-Mills maps
to the strongly coupled string worldsheet theory.
Nevertheless, in some situations elements of 
the YM perturbation theory
can be reproduced from the string theory side.
One of the examples are
the ``spinning strings''.  Spinning strings
are a class of solutions of the classical string worldsheet
theory. They were first considered 
in the context of the AdS/CFT correspondence  
in \cite{FT02,Tseytlin,Russo}. These are strings
rotating in $S^5$ with a large angular momentum.
It was noticed in \cite{FT02} that the
energy of these solutions has an expansion
in some small parameter which is similar in form to
the perturbative expansion in the field
theory on the boundary. Then  \cite{MinahanZarembo}
computed the anomalous dimensions of single
trace operators with the generic large
R-charge, making the actual comparison possible. 
In \cite{FT03} more general solutions were considered,
having large compact charges both in $S^5$ and in $AdS_5$.
For all these solutions,
computations in the classical worldsheet theory
lead to the series in 
the small parameter which on the field theory
side is identified with $\lambda/J^2$ where
$\lambda$ is the 'tHooft coupling constant and
$J$ a large conserved charge. 
Moreover it was shown in \cite{FTQ} that the 
quantum corrections to the classical worldsheet
theory are suppressed for the solutions with the
large conserved charge (see also the recent discussion
in \cite{FPT}). This opened the possibility that
the results of the calculations in the classical mechanics
of spinning strings, which are valid a priori only
in the large $\lambda$ limit, can be in fact extended
to weak coupling and therefore compared to the Yang-Mills
perturbation theory. It was conjectured
 that the Yang-Mills
perturbation theory in the   corresponding sector
is reproduced by the classical dynamics of the spinning
strings. The following picture is emerging.

Single string states
in $AdS_5\times S^5$ correspond to single-trace operators
in the ${\cal N}=4$ supersymmetric Yang-Mills theory. (We consider
the large $N$ limit.)
The dynamics of the single-trace operators 
is described in the perturbation
theory by an integrable spin chain. This spin chain
has a classical continuous limit \cite{Kruczenski}
which describes a class of operators
with the large R-charges. In this limit the
spin chain becomes a classical continuous system.
We have conjectured in \cite{SNS}
that this classical system is equivalent 
to the worldsheet theory of the classical string
in $AdS_5\times S^5$. The Yang-Mills perturbative
expansion corresponds to considering the worldsheet
of the fast moving string as a perturbation of the 
null-surface 
\cite{Kruczenski,SNS,SpeedingStrings,SlowEvolution,KT}.
The null-surface perturbation theory was previously considered
in a closely related context in \cite{dVGN}.

In this paper we will try to make the statement of
equivalence more precise. We will argue that
the string worldsheet theory has a ``hidden'' $U(1)$ symmetry
which is defined unambiguously by its characteristic
properties which we describe. This $U(1)$ commutes
with 
the group of geometrical symmetries of the target space. 
It corresponds to the {\em length of the spin chain} on the
field theory side.
We conjecture that the phase space  of the classical continuous
spin chain is equivalent to the Hamiltonian
reduction of the phase space of the classical
string by the action of this $U(1)$. The equivalence
commutes with the action of geometrical symmetries.

We should stress that the hidden $U(1)$ symmetry which we discuss
in this paper was constructed already
in \cite{KT}, but the explicit calculation
was carried out only at the first nontrivial order
of the null-surface perturbation theory.
The main new result of our paper is that we discuss
this hidden symmetry from the point of view of the integrability.
We conjecture the relation between the $U(1)$ symmetry
and the local conserved charges which if true
gives a uniform description of this symmetry
at all orders of the perturbation theory. 

The classical string on $AdS_5\times S^5$ is an integrable
system (see \cite{MandalSuryanarayanaWadia,BPR,Alday,SwansonMay,SwansonOct} 
and references there), and
our $U(1)$ corresponds to an action variable.
The existence of the action variables for 
integrable systems with a finite-dimensional phase space 
is a consequence of the Liouville theorem
\cite{Arnold}. The classical string has an infinite-dimensional
phase space. We are not aware of the existence
of a general theorem which would
guarantee that the action variables can be constructed in 
the infinite-dimensional case. But we will give two arguments 
for the existence of one action variable for
the string in $AdS_5\times S^5$, at least in the perturbation
theory around the null-surfaces. The first argument gives
an explicit procedure to construct the action variable
order by order in the perturbation theory (Sections 3,  4.4 and 4.6).
The second argument uses the existence of the local
conserved charges \cite{Pohlmeyer} (known as higher Pohlmeyer charges)
and the results of the evaluation of these charges on the
so-called ``rigid solutions''  performed in 
\cite{ArutyunovStaudacher,Engquist}.
The arguments in Section 4 of our paper together with the results of 
\cite{ArutyunovStaudacher,Engquist} suggest that
the action variable is an infinite linear combination
of the Pohlmeyer charges and allow in principle 
to find the coefficients of this linear combination. 

\vspace{10pt}

\noindent
{\em The plan of the paper.}
In Section 2 we will review the classification of
the null-surfaces following mostly \cite{SlowEvolution,SNS} 
and stress that the moduli space
of the null-surfaces is a $U(1)$-bundle over a
loop space. Therefore it has a canonically defined action
of $U(1)$. In Section 3 we will explain how to extend
the action of $U(1)$ from the null-surfaces to the nearly-degenerate 
extremal surfaces using the perturbation theory.
A large part of Section 3 is a
review of \cite{KT}.
In Section 4 we discuss the geometrical meaning of this $U(1)$ 
as an action variable and argue that it
is an infinite sum of the local conserved charges.

\vspace{10pt}
\noindent
{\em Note added in the revised version.}
The coefficients  of the expansion of the action variable in 
the local conserved charges were fixed to all orders in the first
paper of \cite{Later}.
Here we consider only the Pohlmeyer charges
for the $S^5$ part of the string sigma-model.
The role of the Pohlmeyer charges for $AdS_5$ was discussed in
the second paper of \cite{Later}.
In the special  case when the motion of the string
is restricted to ${\bf R}\times S^2\subset AdS_5\times S^5$
the action variable discussed here corresponds to the
action variable of the sine-Gordon model,
see the third paper of \cite{Later}.

\section{Null-surfaces.}\label{sec:NullSurfaces}
\subsection{The definition.}
A  two-dimensional surface in a 
space-time of Lorentzian signature is called a null-surface if it
has a degenerate metric and is ruled by the light rays. 
There is a connection between null-surfaces and
extremal surfaces. An extremal surface is a two-dimensional
surface with the induced metric of the signature $1+1$
which extremizes the area functional. Extremal surfaces
are solutions of the string worldsheet equation of motion
in the purely geometrical background (no $B$-field).
When the string moves very fast, the metric on the worldsheet
degenerates and the worldsheet becomes a null-surface.
Therefore a null-surface can be considered as a degenerate
limit of an extremal surface.

In $AdS_5\times S^5$ there are two types
of the light rays. The light rays
of the first type project to points in $S^5$. 
The light rays of the second type project to the
timelike geodesics in $AdS_5$ and the equator
of $S^5$. The operators of the large R-charge
correspond to the null-surfaces ruled by the
light rays of the second type\footnote{The null-surfaces
of the first type have a boundary. They describe
the shock wave propagating from the cusp of 
the worldline of a spectator quark in 
${\bf R}\times S^3$.}. 

\subsection{The moduli space of null-surfaces.}
\begin{figure} 
\begin{center} 
\epsfxsize=3in {\epsfbox{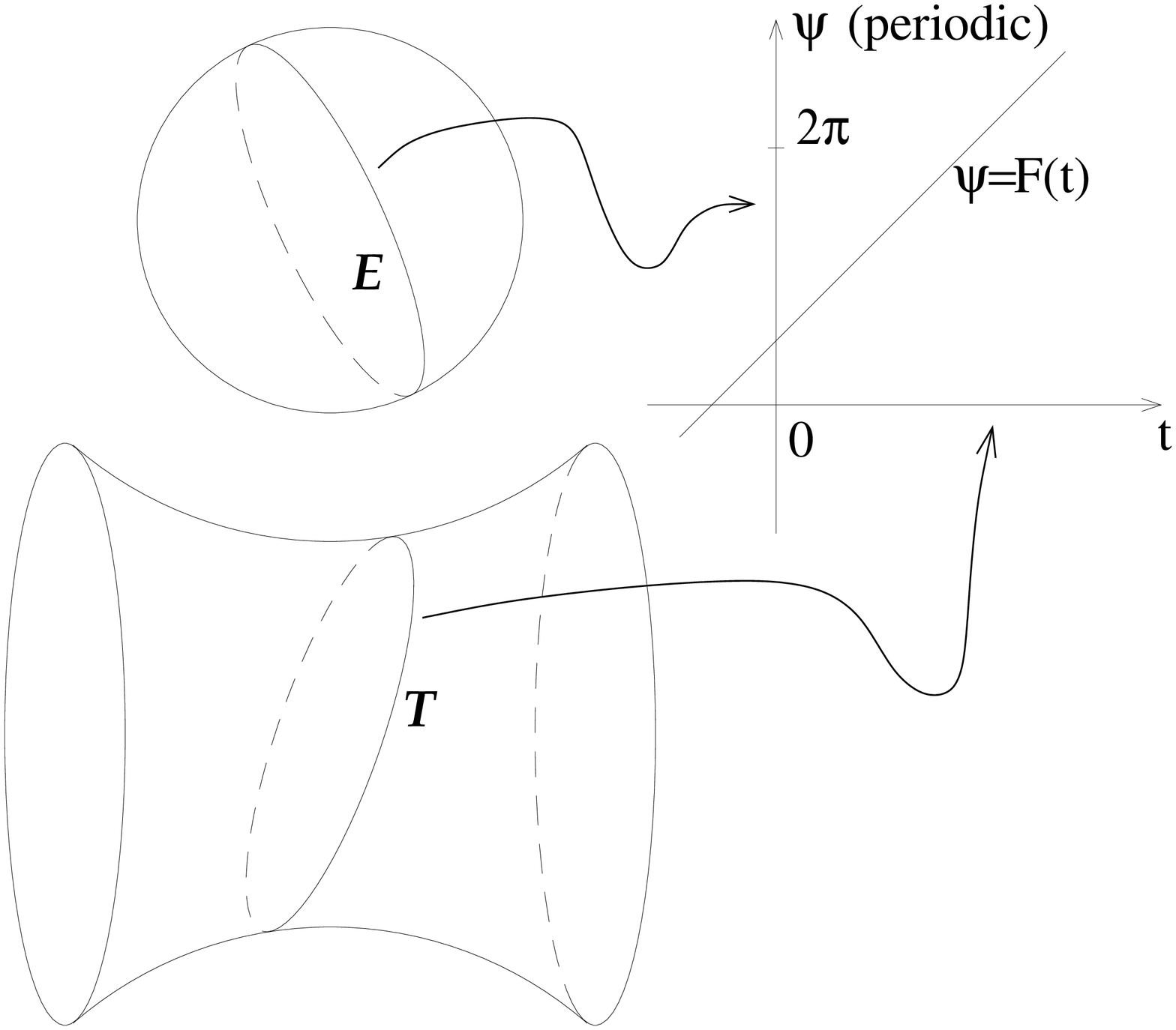}} 
\caption{\small \sl A null-geodesic in $AdS_5\times S^5$ is specified
by the choice of an equator $\bf E$ in $S^5$, a time-like geodesic
$\bf T$ in $AdS_5$ and a map $F:{\bf T}\to {\bf E}$ which maps
the angular parameter $\psi$ on the equator to the time $t$ 
on the geodesic, up to a constant.} 
\label{fig:ng} 
\end{center} 
\end{figure}

It is straightforward to explicitly
describe all the null-surfaces of the second type 
in $AdS_5\times S^5$.
We have
to first describe the moduli space of the null-geodesics
of the second type. 
 An equator of $S^5$
is specified by a point in the
coset space $g_S\in {SO(6)\over SO(2)\times SO(4)}$.
Similarly, a timelike geodesic in $AdS_5$ is specified by
$g_A\in {SO(2,4)\over SO(2)\times SO(4)}$.
Given $g_S$ and $g_A$, let ${\bf E}(g_S)\subset S^5$
and ${\bf T}(g_A)\subset AdS_5$ be the corresponding
equator in $S^5$ and timelike geodesic in $AdS_5$, 
respectively. To specify a light ray in $AdS_5\times S^5$
we have to give also a map $F: {\bf T}\to {\bf E}$
which pulls back the angular coordinate on ${\bf E}$
to the length parameter on ${\bf T}$ (see Fig.\ref{fig:ng}). 
Such  maps
are parametrized by $S^1$. 
We see that each light ray is defined by a triple
$({\bf T},{\bf E},F)$. Therefore, the moduli
space of light-rays of the second type in $AdS_5\times S^5$
is geometrically:
\begin{equation}
\left[{SO(2,4)\over SO(2)\times SO(4)}\times 
{SO(6)\over SO(2)\times SO(4)}\right]\widetilde{\times} S^1
\end{equation}
A null-surface is a one-parameter family of light rays.
Therefore it determines a contour in
$\left[{SO(2,4)\over SO(2)\times SO(4)}\times 
{SO(6)\over SO(2)\times SO(4)}\right]\widetilde{\times} S^1$. 
But we have to also remember
that an arbitrary collection of the light rays is not 
necessarily a null-surface. It is a null-surface only if
the induced metric is degenerate. To understand what it
means, let us 
choose a space-like curve belonging to our surface.
This space-like curve  is a collection of points, 
one point on each light ray. For the surface
to be null, the tangent vector
to this curve at each point of the curve 
should be orthogonal to the
light ray to which the point belongs.
(This condition does not depend on how
we choose a space-like curve.) What kind of a constraint
does it impose on the contour? The space 
$\left[{SO(2,4)\over SO(2)\times SO(4)}\times 
{SO(6)\over SO(2)\times SO(4)}\right]\widetilde{\times} S^1$
is a $U(1)$ bundle over 
${SO(2,4)\over SO(2)\times SO(4)}\times 
{SO(6)\over SO(2)\times SO(4)}$. The condition of
the degeneracy of the metric defines a {\em connection} on
this bundle. The definition of this connection is:
the curve in the total space  is considered
horizontal, precisely if the corresponding collection of
light rays is a degenerate surface. What is the
curvature of this connection? Both 
${SO(2,4)\over SO(2)\times SO(4)}$ and 
${SO(6)\over SO(2)\times SO(4)}$ are Kahler
manifolds (if we forgive that the metric on the first coset 
is not positive-definite). Let us denote the Kahler forms
$k_A$ and $k_S$. The curvature of our $U(1)$-bundle
is $k_A+k_S$. A curve in the base space 
${SO(2,4)\over SO(2)\times SO(4)}\times 
   {SO(6)\over SO(2)\times SO(4)}$ can be lifted to the
   horizontal curve in the total space if and only if
   a two-dimensional film ending on this curve has 
   an integer
   Kahler area (integral of $k_A+k_S$ over this 
   film should be an integer). 
   Moreover, it is lifted as a horizontal curve 
   almost unambiguously, except
   that there is a ``global'' action of $U(1)$
   shifting $F:{\bf T}\to {\bf E}$ on every light ray 
   by the same constant. Therefore, the moduli space
   of null-surfaces is the $U(1)$ bundle
   over the space of contours in
${SO(2,4)\over SO(2)\times SO(4)}\times 
   {SO(6)\over SO(2)\times SO(4)}$ subject to the integrality
   condition which we described.

\begin{figure} 
\begin{center} 
\epsfxsize=5in {\epsfbox{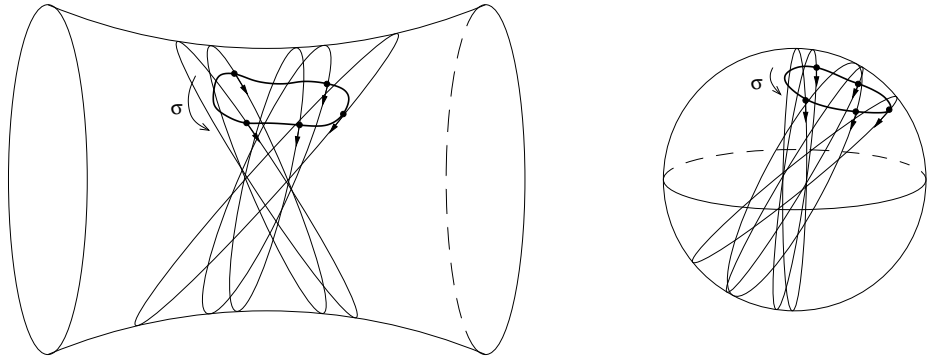}} 
\caption{\small \sl 
A picture of a null-surface in $AdS_5\times S^5$. 
A null-surface is a two-dimensional
surface with the degenerate metric, ruled by the light 
rays. We have shown five light rays
and a spacial contour with a parameter $\sigma$. One can
visualize the null-surface as the surface swept by the 
spacial contour as it moves along the light rays.} 
\label{fig:ns} 
\end{center} 
\end{figure}

To summarize, the moduli space of the null-surfaces
of the second type is:
\begin{equation}\label{ModuliSpace}
	{\mbox{Map}_0\left(S^1,\;\;
{SO(2,4)\over SO(2)\times SO(4)}\times 
{SO(6)\over SO(2)\times SO(4)}\right)
\widetilde{\times} S^1  \over \mbox{Diff}(S^1)}
\end{equation}
Here $\mbox{Map}(S^1,X)$ means the space of maps from the 
circle to $X$; for $X$ a Kahler manifold $Map_0(S^1,X)$ means
the space of maps satisfying the integrality condition.
At this point we consider the null-surfaces without a parametrization;
therefore we divide by the group $\mbox{Diff}(S^1)$
of the diffeomorphisms of the circle.
Turning on the fermionic degrees of freedom on the worldsheet
we get the moduli space of supersymmetric null-surfaces
\cite{SNS}:
\begin{equation}\label{Supersymmetric}
{ \mbox{Map}_0\left(S^1,\;\;
Gr(2|2, 4|4)\right)
\widetilde{\times} S^1  \over \mbox{Diff}(S^1)}
\end{equation}
Here $\mbox{Map}_0\left(S^1,\;\;
Gr(2|2, 4|4)\right)$ is the phase space of the continuous
spin chain \cite{SNS}. Therefore the moduli space of null-surfaces
is ``almost'' equivalent to the phase space of the continuous
spin chain, except for the fiber $S^1$ and the reparametrizations
$\mbox{Diff}(S^1)$.  We have to explain
what happens to the fiber and why the null-surface
actually comes with the parametrization. 
Also, we have to explain how the symplectic structure
is defined on the moduli space of null surfaces.
Let us start with the parametrization.

\subsection{Parametrized null-surfaces.}\label{sec:Parametrization}
The phase space of the classical string has a boundary which
consists of strings ``moving with the speed of light''.
A string moving very fast can be approximated by a null-surface.
But one null-surface can approximate many different
fast moving strings. The null-surface as we defined
it so far ``remembers'' only the {\em direction} of the velocity
at each point of the approximated string, but it misses the information
about the ratios of the relativistic factors $\sqrt{1-v^2}$
at different points of the string. Although $\sqrt{1-v^2}\to 0$
in the null-surface limit, the ratio ${\sqrt{1-v^2(\tau,\sigma_1)}/
\sqrt{1-v^2(\tau,\sigma_2)}}$ for two different points on
the worldsheet remains finite. 
Therefore, if we want to think of the moduli space
of the null-surfaces as the boundary of the phase space,
we have to equip the null-surfaces with an additional structure.
This additional structure is the  {\em parametrization}.

A null-surface is a one-parameter family of
the light rays. The parameterization is a particular choice of
the parameter. In other words, it is a monotonic function $\sigma$ from
the family of  light rays forming the null-surface to the circle, defined
modulo $\sigma\sim\sigma+\mbox{const}$.
One can also think of it as a density $d\sigma$ on the
set of light rays forming the null-surface. 
This density is roughly speaking proportional to 
the density of energy on the worldsheet of the fast-moving string,
in the limit when it becomes the null-surface.
We will now give the definition of $\sigma$.

Consider the family of string worldsheets $\Sigma(L)$
converging  to the null-surface 
$\Sigma_0=\Sigma(\infty)$. 
We will introduce a parametrization $d\sigma$ of $\Sigma_0$ 
in the following way.
 Consider a Killing vector field $U$ on
$S^5$, corresponding to some rotation of the sphere: 
\begin{equation}
	U.x_S^i=u^{ij}x_S^j
\end{equation}
Here $x_S^i$ parametrizes the $S^5$: $\sum (x_S^i)^2=1$.

When $L$ is large, $\Sigma(L)$ is close to $\Sigma_0$, the string
moves very fast and the conserved charge corresponding to 
$U$ is very large. We can approximate this charge by an integral
over a spacial contour on the null-surface $\Sigma_0$ of 
$u^{ij} x_{0,S}^i\partial_{\tau}x_{0,S}^j$ times some density
$d\sigma$:
\begin{equation}\label{NCD}
	Q_U=L\int_{\sigma\in [0,2\pi]} d\sigma\; u^{ij} \;
	x_{0,S}^i\partial_{\tau}x_{0,S}^j+(\mbox{terms vanishing at }L\to \infty)
\end{equation}
Here $x_{0,S}$ is the $S^5$-part of the null-surface; 
we choose the $\tau$ coordinate on the null-surface to be
the affine parameter on the light ray normalized by the condition
$x_{0,S}(\tau+2\pi ,\sigma)=x_{0,S}(\tau,\sigma)$.
Eq. (\ref{NCD}) with the condition
 $\int d\sigma =2\pi$ is the definition of $d\sigma$, and
 also the precise definition of the large parameter $L$, modulo
 $O(1/L)$.
 We choose $\sigma$ as the parametrization.
 
We can now say that {\em the moduli space
of parametrized null-surfaces is the boundary
of the phase space of a classical string.} 
We say that a family $\Sigma(L)$ of extremal
surfaces has a {\em parametrized} null-surface $\Sigma_0$
as a limit when $L\to\infty$ if and only if
\begin{itemize}
	\item 
$\Sigma(L)$ has $\Sigma_0$ as a limit when $L\to\infty$,
as a continuous family of smooth two-dimensional surfaces
in a smooth two-dimensional manifold, and 
 \item the density of $Q_U$ approaches Eq. (\ref{NCD})
	 in the limit $L\to\infty$.
\end{itemize}
This definition of the parametrization does not
depend on which particular geometrical symmetry $U$ we use. 
An alternative way to define the same parametrization is to use
a special choice of 
the worldsheet coordinates on $\Sigma$. Let us choose the
worldsheet coordinates $\tau,\sigma'$ so that 
\begin{eqnarray}
&&	\left({\partial x_S\over\partial \tau}\right)^2+
	\left({\partial x_S\over\partial \sigma'}\right)^2
	=-\left({\partial x_A\over\partial \tau}\right)^2-
	\left({\partial x_A\over\partial \sigma'}\right)^2
	=1 \nonumber \\
&&	\left({\partial x_S\over\partial \tau},
	{\partial x_S\over\partial \sigma'}\right)=
	-\left({\partial x_A\over\partial \tau},
	{\partial x_A\over\partial \sigma'}\right)=\mbox{const}
	\nonumber
\end{eqnarray}
where $x_A$ is the projection of the string worldsheet
to $AdS_5$ and $x_S$ is the projection to $S^5$.
Then we define $\sigma=\sigma'/\int d\sigma'$. In the null-surface
limit $d\sigma$ defines the parametrization of the null-surface.

\subsection{The symplectic structure.}
The moduli space of parametrized null-surfaces as a manifold
depends only on the conformal structure of the target space.
But we can introduce additional structures on this moduli
space which use the metric on $AdS_5\times S^5$.

An important additional structure is the closed 2-form
which originates from the symplectic form of the classical string.
Strictly speaking a differential form in the bulk
of the manifold does not automatically determine a differential
form on the boundary. Indeed, suppose that we have
a differential form, for example a 2-form $\Omega$ in the bulk.
We can try to define the ``boundary value'' $\omega$ of $\Omega$
on the boundary in the following way. Given two vector
fields $v_1,v_2$ on the boundary, we find two vector
fields $V_1,V_2$ in the bulk such that $\lim V_1=v_1$
and $\lim V_2=v_2$. Then we define $\omega(v_1,v_2)=\lim\Omega(V_1,V_2)$.
But the problem is that this definition will depend on the
choice of $V_1$ and $V_2$. Intuitively, if $(\tilde{V}_1,\tilde{V}_2)$ 
is some other choice of a pair of vector fields inducing
$(v_1,v_2)$ on the boundary, and the ``vertical component''
of $\tilde{V}_i-V_i$ is not small enough near the boundary,
then $\Omega(V_1,V_2)\neq \Omega(\tilde{V}_1,\tilde{V}_2)$.

Given this difficulty, how do we define the symplectic
form on the space of null-surfaces given the
symplectic form on the string phase space?
When we lift the vector field
$v$ on the boundary to the vector field $V$ in the
bulk, let us require that $dL(V)$ goes to zero when $L\to\infty$. 
We define $L$  by Eq. (\ref{NCD}); it is only an approximate
definition at $L\to\infty$, but this is good enough for the purpose of
our definition:
\begin{equation}
	\omega(v_1,v_2)=\lim_{L\to\infty} L^{-1} \Omega(V_1,V_2)
\end{equation}
where $V_1$ and $V_2$ are such that 
$dL(V_1)=dL(V_2)\simeq 0$.
One can see that $\omega$ has a kernel, which is
precisely the tangent space to the fiber $S^1$ 
in the numerator of Eq. (\ref{Supersymmetric}).
The moduli space has a symmetry $U(1)$ rotating this fiber;
we will discuss this symmetry in the next section;
we will call it $U(1)_L$.
Therefore $\omega$ is the symplectic form on
the moduli space of null-surfaces modulo $U(1)_L$.

Eq. (\ref{Supersymmetric}) implies that the moduli
space of {\em parametrized} null-surfaces {\em modulo}
$U(1)_L$ is the space of parametrized contours
in the Grassmanian:
\begin{equation}
	\mbox{Map}_0\left(S^1,\;\; Gr(2|2, 4|4)\right)
\end{equation}
One can see that $\omega$ is equal to the integral of the symplectic form
on the super-Grassmanian pointwise on the contour, with the measure $d\sigma$.
The symplectic area of the film filling the contour
is the generating function of the shift of the origin
of the circle. Therefore the integrality condition
guarantees that the symplectic form does not depend
on the choice of the origin on $S^1$; the symplectic form
is horizontal and invariant with respect to the shifts
of the origin of $S^1$. 

Our definition of the symplectic form
on the space of null-surfaces used the target space metric
 (just the conformal structure would not be
enough) and also the fact that the target
space is a product of two manifolds.

\section{Nearly-degenerate extremal surfaces and 
the role of the engineering dimension.}
Our discussion in this and the next section will be limited to 
the  classical  bosonic string.
\subsection{Definition of $U(1)_L$.}\label{ChargeDefinition}
\begin{figure}
\begin{center} 
\epsfxsize=2in {\epsfbox{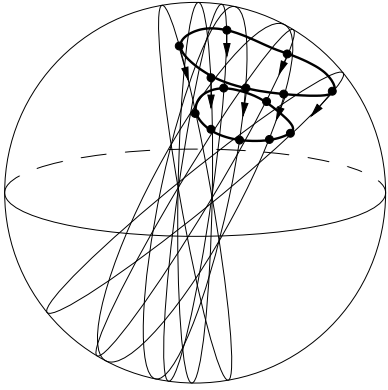}} 
\caption{\small \sl 
The action of $U(1)_L$ on the null-surface.
The symmetry acts only on the $S^5$-part of the null-string.
Each point shifts by the same angle along the
equator which is the projection to $S^5$ of the
corresponding light ray.} 
\label{fig:nsS} 
\end{center} 
\end{figure}

The moduli space (\ref{Supersymmetric}) of null-surfaces
is a $U(1)$-bundle. The $U(1)$ symmetry shifting in
the fiber $S^1$
plays an important role in the formalism. 
We will call it $U(1)_L$. On fig. \ref{fig:nsS} we have shown 
schematically how
$U(1)_L$ acts on the null-surfaces.
We conjecture
that $U(1)_L$ corresponds to the length of the spin chain.
Generally speaking, the length of the spin chain is 
not conserved in the Yang-Mills perturbation 
theory \cite{BeisertDynamic}, 
but it is probably conserved in the
continuous limit (this should be related
to the discussion of the ``closed  sectors'' in 
\cite{Minahan}). It should be conserved modulo the
corrections vanishing in the continuous
limit. 
We therefore conjecture that there is a continuation of 
$U(1)_L$ from the space of null-surfaces to the phase space
of the classical string, at least to the region of the phase
space corresponding to fast moving strings. 
We conjecture that this continuation is uniquely defined by the
following properties: 
\begin{enumerate}
	\item The action of $U(1)_L$ preserves the symplectic
		structure.
	\item The action of $U(1)_L$ does not change the
		projection of the worldsheet to $AdS_5$.
		Moreover, it preserves the projection to
		$AdS_5$ of the null-directions on the
		worldsheet.
	\item We require that the orbits
		of $U(1)_L$ are closed (otherwise, we would 
		not have called it $U(1)$).
	\item The restriction of $U(1)_L$ to the null-surfaces
		acts as we described (see fig. \ref{fig:nsS}).
\end{enumerate}
The second property reflects the fact that $U(1)_L$ corresponds
to the length of the operator rather than its engineering 
dimension.

Let ${\cal E}$ denote the Hamiltonian 
of $U(1)_L$. Let $X$ denote the phase space of the classical
string, and $X//({\cal E}=l)$ denote the Hamiltonian
reduction of the phase space on the level set of $\cal E$. 
The basic conjecture is:
\begin{quote}
There is a one-to-one map from the phase space of the
spin chain of the length $l$ to the reduced phase space of the
classical string $X//({\cal E}=l)$ preserving the 
symplectic structure and commuting with the action of 
$SO(2,4)\times SO(6)$.
\end{quote}
The reduction by $U(1)_L$ was discussed in \cite{KMMZ}
but only in a sector \cite{BeisertSector} in which
$U(1)_L$ acts as some element of $SO(6)$. The perturbation
theory in this sector was discussed in \cite{KRT} (see also
Section 2 of \cite{SlowEvolution}).

\subsection{Action of $U(1)_L$ on nearly-degenerate
extremal surfaces.}
In this subsection we will explain how to continue
the action of $U(1)_L$ from the boundary of the phase
space.
Most of this section is  a partial 
review\footnote{Section 3 of \cite{KT} has more
than just a construction of $U(1)_L$. 
The next step is considering the action of the
Killing vector field
$\partial\over\partial T$ where $T$ is the global time
in $AdS_5$ on the invariants of $U(1)_L$ and bringing
the result to the form suitable for the comparison
with the field theory computation. Here we are discussing
only the first step.}
of Section 3 of \cite{KT}.

\subsubsection{Particle on a sphere.}\label{CaseOfParticle}
Consider the phase space of a particle moving on $S^5$,
and restrict to the domain where the velocity of 
the particle is nonzero.
This domain is naturally 
a bundle over the moduli space of equators of $S^5$;
let $\pi$ denote the projection map in this bundle.
A point of the phase space, corresponding
to the position  $x\in S^5$ and the velocity 
$v\in T_x S^5$, projects by $\pi$ to the equator
going through $x$ and tangent to $v$. 
See the discussion in \cite{SlowEvolution}. 

The symplectic form on the phase space is expressed
in terms of the symplectic form on the base
and the connection form ${\cal D}\psi$:
\begin{equation}
	\omega=df\wedge {\cal D}\psi+
	f\pi^*\Omega
\end{equation}
where ${\cal D}\psi={(p,dx)\over \sqrt{(p,p)}}$,
$f=\sqrt{(p,p)}$ ($p$ is the momentum of the particle)
and $\Omega$ is the symplectic form
on the moduli space of equators. The moduli space of equators
$SO(6)\over SO(2)\times SO(4)$ is a Kahler manifold,
the symplectic form $\Omega$ is the Kahler form.

Now it is easy to construct the action of $U(1)$.
One takes
\begin{equation}\label{ParticleUOne}
	{\cal V}={\partial\over\partial\psi}
\end{equation}
This is a vertical vector field, it does not act on the
base. The coordinate
$\psi$ is essentially the angle along the equator on which the
particle is moving. More explicitly:
\begin{equation}\label{ExplicitDPsi}
	{\partial\over\partial \psi}.x=
	{1\over \sqrt{(\partial_{\tau}x,\partial_{\tau}x)}}\;
	\partial_{\tau}x
\end{equation}
It is easy to see that the trajectories of the vector
field $\partial\over\partial\psi$ on the phase space of
a particle on $S^5$ are periodic with the period $2\pi$.
One has to remember that this vector field is defined
only on the open subset of the phase space, where the
velocity of the particle is nonzero. But we consider
fast moving strings, and the region of the phase
space where the velocity is nearly zero is not
important for  us. 

\subsubsection{String on a sphere.}\label{StringOnASphere}
In some sense, a string is a continuous collection of particles.
Therefore, it is natural to apply a similar construction
to the string. Treating the string as a continuous collection
of particles requires the choice of the coordinates on
the worldsheet. We will therefore introduce the conformal gauge:
\begin{equation}\label{ConformalGauge}
	\begin{array}{l}
	(\partial_{\tau}x)^2+(\partial_{\sigma}x)^2=0\\[5pt]
	(\partial_{\tau}x,\partial_{\sigma}x)=0
\end{array}
\end{equation}
In this gauge the symplectic form is:
\begin{equation}\label{omega}
	\omega=\oint d\sigma
	(\delta_1x,\;\stackrel{\leftrightarrow}{D}_{\tau}
	\delta_2x)
\end{equation}
  In the Hamiltonian formalism, we introduce 
$p_A=\partial_{\tau}x_A\in T(AdS_5)$ --- the $AdS_5$-component
of the momentum, and $p_S=\partial_{\tau}x_S\in T(S^5)$ 
--- the $S^5$-component of the momentum.
Now we will interpret the string as a collection
of particles parametrized by $\sigma$. We are tempted
to interpret the vector field 
(\ref{ParticleUOne}),(\ref{ExplicitDPsi})
acting pointwise in $\sigma$ as  the required 
$U(1)_L$ symmetry. 
The generator of this symmetry would be $\int d\sigma |p_S|$.
But this would be wrong. This field preserves
the symplectic structure, 
does have periodic trajectories and acts correctly on the
null surfaces. But unfortunately
it does not preserve the gauge (\ref{ConformalGauge}).
It only commutes
with the second constraint, 
$(p,\partial_{\sigma}x)=0$.
But it does not commute with the first one,
$(p,p)+(\partial_{\sigma}x,\partial_{\sigma}x)=0$.
Indeed,
it commutes with $(\partial_{\tau}x_S)^2=(p_S,p_S)$
but not with $(\partial_{\sigma}x_S)^2$.
Therefore we should modify this vector field 
so that it still has periodic trajectories, but also
commutes with the constraint. There is a systematic
procedure to do this, order by order in $1\over (p_S,p_S)$,
developed in \cite{KT}.

Let us summarize this procedure, or perhaps a variation of it.
To make sure that the modified vector field is Hamiltonian
(preserves the symplectic structure) we construct it as
a conjugation of $\partial\over\partial \psi$ with 
some canonical transformation, which we denote $F$:
\begin{equation}
	{\cal V}.x=F^{-1}\left[{\partial\over\partial \psi}.
	F[x]\right]
\end{equation}
or schematically ${\cal V}=F^{-1}\circ {\partial\over\partial\psi}\circ F$.
Since $F$ is a canonical transformation, ${\cal V}$ is automatically
a Hamiltonian vector field. 
Since $F$ is single-valued, $\cal V$
generates periodic trajectories. It remains to
construct $F$ such that $\cal V$ commutes with the
constraint $(p,p)+(\partial_{\sigma}x)^2$.
But to require that 
$F^{-1}\circ {\partial\over\partial \psi}\circ F^{-1}$ 
commutes with 
$(p,p)+(\partial_{\sigma}x)^2$ is the same
as to require that $\partial\over\partial\psi$
commutes with $F^*[(p_S,p_S)+(\partial_{\sigma}x_S)^2]$ ---
the pullback of $(p_S,p_S)+(\partial_{\sigma}x_S)^2$ by $F$.
Therefore we have to find such a canonical transformation $F$
that the pullback of $(p_S,p_S)+(\partial_{\sigma}x_S)^2$
with $F$ is annihilated by the vector field 
$\partial\over\partial\psi$. In other words, we
have to find a canonical transformation which
removes $\psi$ from $(p_S,p_S)+(\partial_{\sigma}x_S)^2$;
after this canonical transformation 
$|p_S|^2+(\partial_{\sigma}x_S)^2$ becomes 
$|p_S|^2+\phi_0+\phi_1+\ldots$
where all the $\phi_k$ for $k\geq 0$ are in involution
with $\int |p_S(\sigma)| d\sigma$ and $\phi_k$ is of
the order $1/|p_S|^{2k}$.
This was done in Section 3 of \cite{KT}. 
The canonical transformation can be expanded
in $1/(p_S,p_S)$; the corresponding generating function
is expanded in the odd powers of $1/|p_S|$.
The authors of \cite{KT} gave the explicit
expression for $F$ to the first order in $1/|p_S|$, 
but
they also give a straightforward algorithm for constructing
the higher orders. (We will reconsider the higher orders
from a slightly different point of view 
in Section \ref{sec:Alternative}, perhaps
making this algorithm more precise.)

At the first order
we need to find $h^{(1)}$ such that the canonically
transformed constraint, which
is a function of $\sigma$:
$$
	(p_S,p_S)(\sigma)+
	(\partial_{\sigma}x_S,\partial_{\sigma}x_S)(\sigma)+
	\{h^{(1)},\; \left[(p_S,p_S)(\sigma)+
	(\partial_{\sigma}x_S,\partial_{\sigma}x_S)(\sigma)
	\right]\}
$$
has zero Poisson bracket with $\int d\sigma' |p_S|(\sigma')$ 
up to  the terms of the order $1/|p_S|^3$, for every $\sigma$.
And $h^{(1)}$ should be of the order $1/|p_S|$. In other
words, we should have:
\begin{equation}\label{ForAnySigma}
	\left\{\int d\sigma |p_S|(\sigma), \; 
	\left((\partial_{\sigma'}x_S)^2(\sigma')
	+\{h^{(1)},\;|p_S|^2(\sigma')\}\right)\right\}=0
\end{equation}
One can  see that 
\begin{equation}
	h^{(1)}=-{1\over 4}\int {d\sigma \over |p_S|}
	\left(\partial_{\sigma}x_S,\;D_{\sigma}{p_S\over |p_S|}
\right)(\sigma)
\end{equation}
works. Notice also that this $h^{(1)}$ is reparametrization
invariant 
(where $|p_S|$ transforms as a density of weight one).
Therefore it commutes also with the second constraint
$(p_S,\partial_{\sigma}x_S)=\mbox{const}$.
Therefore, to the first order in $1/|p_S|$ the
canonical transformation we are looking for  is generated by
this $h^{(1)}$. Then the generator of the $U(1)_L$
is, up to the terms of the order $1/|p_S|^3$:
\begin{eqnarray}\label{PeriodicSecondOrder}
	&& {\cal E}=\int d\sigma |p_S|-
	\{h^{(1)},\; \int d\sigma |p_S|(\sigma)\}
	+\ldots =\\[10pt]&&
	=\int d\sigma \left[|p_S|+
	{1\over 4|p_S|}
\left[(\partial_{\sigma}x_S)^2-
\left(D_{\sigma}{p_S\over |p_S|},D_{\sigma}{p_S\over |p_S|}
\right)-
{(p_S,\partial_{\sigma}x_S)^2\over (p_S,p_S)}\right]
+\ldots\right]
\nonumber
\end{eqnarray}
One can see immediately that the trajectories of this
charge are closed up to the terms of the order
subleading to $1/|p_S|$. Indeed, we have explained
in Section \ref{CaseOfParticle} why the leading
term gives periodic trajectories. And the second term
(which as we have seen is needed to make the charge
commuting with the Virasoro constraints) averages
to zero on the periodic trajectories of the
first term. Therefore
(see for example Section 3 of
\cite{SlowEvolution}) the trajectories of ${\cal E}$ 
do not drift at this order.

We will discuss the higher orders
in Section \ref{sec:Alternative}.

\section{Length of the operator 
and local conserved charges.}\label{sec:SectionFour}
We have seen that the null-surface perturbation theory
has a ``hidden'' symmetry $U(1)_L$.
The existence of  $U(1)$ symmetries acting on the phase
space is typical for integrable systems, at least
for those which have a finite-dimensional phase space. 
Corresponding
conserved charges are called action variables \cite{Arnold}.
Classical string in $AdS_5\times S^5$ is an integrable
system.
Therefore, we should not be surprised to find
such an action variable\footnote{Strictly speaking,
the integrability is not necessary for the construction
of the action variable {\em in perturbation theory}.
A typical example is a particle on a sphere $S^2$ in 
an arbitrary (polynomial) potential. When the particle
moves very fast, it does not feel the potential. All the
trajectories are periodic in the limit of an infinite velocity. 
Therefore on the boundary of the phase space, when the 
velocity is infinite, we have
an action variable $|p|$ --- the absolute value of the
momentum. It is well known that the 
perturbation theory in $1/|p|$ allows us to extend
this action variable from the boundary inside the phase space,
but {\em only in the perturbation theory}. For an arbitrary
potential, the perturbation series must diverge, because
in fact there is no additional conserved quantity besides the energy. 
Therefore the $U(1)$ will be actually broken by effects which
are not visible in the perturbation theory, unless if 
the potential is such that the system is integrable.
We want to thank V.~Kaloshin and A.~Starinets for 
discussions of this subject.}.

The local conserved charges in involution for the classical string
in $AdS_5\times S^5$ are explicitly known. Therefore,
instead of constructing $U(1)_L$ in the perturbation theory,
we can try to build it as some linear combination
of the already known conserved charges.  
In this section, we will argue that the coefficients
of this linear combination are actually fixed
by the calculation of \cite{ArutyunovStaudacher,Engquist}.

\subsection{Local conserved charges.}\label{exact}
Consider a string in the target space which is a product of two
manifolds $A$ and $S$. We assume that the metric on $A$
has the Lorentzian signature, and the metric on $S$
has the Euclidean signature.
We will need $A=AdS_5$ and $S=S^5$,
but let us first consider the general $A\times S$.
The string worldsheet will be denoted $\Sigma$.
The classical trajectory of the string is an embedding
$x:\Sigma\to A\times S$. We are going to use the
fact that the target space is a direct product.
A point of $A\times S$ is obviously a pair $(x_A, x_S)$
where $x_A$ is a point of $A$ and $x_S$ is a point of $S$. 
Therefore for each point $\zeta\in\Sigma$ we have
$x(\zeta)=(x_A(\zeta),x_S(\zeta))$, where
$x_A\in A$ and $x_S\in S$. 
Consider the 1-forms $dx_A$ and $dx_S$ 
on the string worldvolume, $dx_A$ taking values
in $T_x A$ and $dx_S$ in $T_x S$. In other words,
$\left[\begin{array}{c}dx_A\\ dx_S\end{array}\right]$ 
	is a differential of $x$. 

The metric on $A\times S$ has the Lorentzian
signature, and we consider the string worldsheets
which have the induced metric with the Lorentzian
signature. Pick two vector fields $\xi_+$ and $\xi_-$ 
on $\Sigma$,
which are both lightlike but have a nonzero scalar product:
\begin{eqnarray*}
 (\xi_+,\xi_+)=0 \\
 (\xi_-,\xi_-)=0\\
 (\xi_+,\xi_-)\neq 0
\end{eqnarray*}
These vector fields have a simple geometrical meaning.
Since the worldsheet is two-dimensional, at each point
we have two different lightlike directions. The vector
$\xi_+$ points along one lightlike direction, 
and $\xi_-$ points along another. 
Pick a spacial
contour $C$ on $\Sigma$, and a 1-form $\nu$ on $\Sigma$
such that $\nu(\xi_-)=0$ and $\nu(\xi_+)\neq 0$.
Consider the following functional:
\begin{equation}\label{ChargeXiNu}
	Q^{[1]}[x]=
	\oint_C \nu {\sqrt{(dx_S(\xi_+))^2}\over \nu(\xi_+)}
\end{equation}
We will prove that this functional 
does not depend on a particular
choice of $\xi_+$, $\xi_-$, $\nu$ and $C$.
This is therefore a correctly defined functional on the
phase space of the string. Indeed, the only ambiguity
in the choice of $\xi_+$ is 
$\widetilde{\xi}_+=f(\zeta)\xi_+$
where $f$ is some function on the worldsheet. But this
function cancels in (\ref{ChargeXiNu}).  The ambiguity in
the choice of $\xi_-$ and $\nu$ is also in rescaling
which does not change (\ref{ChargeXiNu}).
It remains to prove that (\ref{ChargeXiNu}) does not depend
on the choice of the integration contour $C$.
To prove that (\ref{ChargeXiNu}) is independent of $C$,
let us choose coordinates $(\tau^+,\tau^-)$ 
on the worldsheet in such a way
that the induced metric is 
$ds^2=\rho(\tau^+,\tau^-)d\tau^+ d\tau^-$.
Then $\xi_+$ is proportional to $\partial\over\partial \tau^+$
and $\xi_-$ is proportional to $\partial\over\partial \tau^-$.
In these coordinates
\begin{equation}\label{QOne}
	Q^{[1]}=\oint_C  d\tau^+ \sqrt{(\partial_{+}x_S)^2}
\end{equation}
The variation of $Q^{[1]}$ under the variation of the contour
is measured by the differential of the form:
\begin{eqnarray*}
	d\left(d\tau^+ \sqrt{(\partial_{+}x_S)^2}\right)=\\
	=-d\tau^+\wedge d\tau^- {(\partial_{+}x_S,
	D_-\partial_+x_S)\over \sqrt{(\partial_{+}x_S)^2}}
\end{eqnarray*}
But on the equations of motion $D_+\partial_-x_S=0$.
Therefore the integral does not depend on the choice of the
contour. 

Let us explain why on the equations of motion we have
$D_+\partial_-x_S=0$. Let $N$ be the second quadratic form 
of the surface, $N: S^2(T\Sigma)\to {\cal N}\Sigma$ 
(here ${\cal N}\Sigma=T(A\times S)/T\Sigma$ is the normal
bundle to $\Sigma$ in $A\times S$). The second quadratic form
is defined in the following way: suppose that the particle
moves on $\Sigma$ with the velocity $v$, then the acceleration
of the particle is $N(v)$ modulo a vector parallel to 
$T\Sigma$. For the surface to be extremal, the trace of
$N$ should be zero. The trace of $N$ is the contraction of
$N$ with the induced metric on $\Sigma$; it is a section
of ${\cal N}\Sigma$. The trace of $N$ is proportional to
$D_+\partial_-x$, therefore
we should have:
$$
D_+\partial_-x=f^+(\tau^+,\tau^-)\partial_+x+
f^-(\tau^+,\tau^-)\partial_-x
$$
But notice that $(D_+\partial_-x,\partial_-x)=
(D_-\partial_+x,\partial_+x)=0$ therefore
$f^+=f^-=0$.
Another conserved charge is:
\begin{equation}\label{QOneTilde}
	\widetilde{Q}^{[1]}=
	\oint_C  d\tau^- \sqrt{(\partial_{-}x_S)^2}
\end{equation}
Are there charges containing higher derivatives of $x_S$?
Let us consider the following expression:
\begin{equation}
	J_+^{[2]}(\tau_+,\tau_-)={1\over |\partial_+x_S|}
	\left(D_+{\partial_+ x_S\over |\partial_+x_S|},
	D_+{\partial_+ x_S\over |\partial_+x_S|}\right)
\end{equation}
Even though $D_+\partial_-x_S=0$ 
it is not true that $\partial_- J_+^{[2]}$ is zero.
The covariant derivatives $D_+$ and $D_-$ do not
commute, therefore $D_-D_+{\partial_+ x_S\over 
|\partial_+ x_S|}\neq 0$.
In fact, for any function $w:\Sigma\to T(A\times S)$
we have
\begin{equation}
	[D_+,D_-]w=R(\partial_+x,\partial_-x).w
\end{equation}
where $R$ is the Riemann tensor of $A\times S$.
Now we have to start using that 
$S$ is a sphere.
For $S=S^5$, the Riemann tensor is constructed from
the metric tensor, and
\begin{equation}\label{SpecialRiemann}
	[D_+,D_-]w=\partial_+x (\partial_-x,w)-
	\partial_-x (\partial_+x, w)
\end{equation}
Now consider the following differential form:
\begin{equation}\label{ChargeUV}
	\lambda= 2{d\tau^- \over |\partial_+x_S|}
	(\partial_-x_S,\partial_+x_S)+
	{d\tau^+\over |\partial_+x_S|}
	\left(D_+{\partial_+ x_S\over |\partial_+x_S|},
	D_+{\partial_+ x_S\over |\partial_+x_S|}\right)
\end{equation}
Using (\ref{SpecialRiemann}) we can show that
$d\lambda=0$, therefore $\oint \lambda$
is a local conservation law. We use the formula
$D_-D_+\partial_+x_S=(\partial_+x_S)^2\partial_{-}x_S
-(\partial_+x_S,\partial_-x_S)\partial_+x_S$
which is special for $S^5$. We will denote this charge
$Q^{[2]}$. There is also a charge $\widetilde{Q}^{[2}$
which is obtained from (\ref{ChargeUV})
by replacing $\tau^+$ with $\tau^-$ and $\partial_+$ or $D_+$
with $\partial_-$ or $D_-$.

These charges are just the first examples of an infinite
family of charges, which are all in involution.
This infinite family was constructed in \cite{Pohlmeyer}.

A particularly important linear combination is
\begin{equation}\label{E2}
{\cal E}_2={1\over 2}(Q^{[1]}-\tilde{Q}^{[1]})
\end{equation}
The construction of this 
charge requires only that the target space is a direct product
of two manifolds. 

\subsection{Local conserved charges are invariant
under $U(1)_L$.}\label{sec:Commutes}
Consider a local conserved charge
$Q$ acting trivially on the $AdS$ part of the worldsheet.
In the conformal gauge, this means that $Q$ is constructed
as a contour integral of some combination of $x_S$ and $p_S$.
Let us decompose $Q$ in the inverse powers of $|p_S|$:
\begin{equation}\label{QExpansion}
	Q=Q_m+Q_{m+1}+Q_{m+2}+\ldots
\end{equation}
where $m$ is a non-negative integer, the ``order'' of
the charge; $Q_m$ is of the order $1/ |p_S|^{2m-1}$,
$Q_{m+1}$ is of the order $1/ |p_S|^{2m+1}$ etc.
We have to require that $Q$ is in involution with
the Virasoro constraints. 
In particular, it should be in
involution with $|p_S(\sigma)|^2+
(\partial_{\sigma}x_S(\sigma))^2$ for an arbitrary 
$\sigma$. 
(Here we used that $Q$ is trivial in AdS-part.)
Let us now apply the canonical transformation
$F$ which we described in Section \ref{StringOnASphere}.
After this canonical transformation 
$|p_S|^2+(\partial_{\sigma}x_S)^2$ becomes 
$|p_S|^2+\phi_0+\phi_1+\ldots$
where all the $\phi_k$ for $k\geq 0$ are in involution
with $\int |p_S(\sigma)| d\sigma$ and $\phi_k$ is of
the order $1/|p_S|^{2k}$. And $Q=Q_m+Q_{m+1}+\ldots$
becomes $Q'=Q'_m+Q'_{m+1}+\ldots$, where $Q'$ is the
canonically transformed $Q$. 
We should have:
\begin{equation}
	\{ |p_S(\sigma)|^2+\phi_0(\sigma)+\phi_1(\sigma)
	+\ldots ,\;\;
	Q'_m + Q'_{m+1}+\ldots \}=0
\end{equation}
for an arbitrary $\sigma$.
At the leading order in $|p_S|$ this implies that
	$\int d\sigma |p_S(\sigma)|$ is in involution
	with $Q_m'$. At the next order, it follows
	that for all values of $\sigma$ the expression 
	$\{ |p_S(\sigma)|^2, Q'_{m+1} \}$ is in involution
	with $\int d\sigma' |p_S(\sigma')|$. This implies
	that:
\begin{equation}
\left\{ \int d\sigma' |p_S(\sigma')| ,
\left\{ \int d\sigma |p_S(\sigma)| ,\;
Q'_{m+1}\right\}\right\}=0
\end{equation}
Since the vector field generated by $\int d\sigma |p_S(\sigma)|$
is periodic, this equation implies that
$\int d\sigma |p_S(\sigma)|$ is in involution with $Q_{m+1}'$.
An analogous argument at higher orders shows that all the
$Q'_{m+j}$ commute with $\int d\sigma |p_S(\sigma)|$.
Therefore $Q'$ is in involution  with 
the expression $\int d\sigma |p_S(\sigma)|$ which is the
generator of $U(1)_L$. 
The conserved charges of \cite{Pohlmeyer} do have an expansion
of the form (\ref{QExpansion}) therefore they should
commute with $U(1)_L$. This reinforces our conjecture
that $U(1)_L$ should be a combination of the local conserved
charges.

\subsection{A geometrical meaning 
of $U(1)_L$.}\label{Liouville}
We can try to make more transparent the geometrical meaning of
$U(1)_L$ by drawing an analogy with the Liouville theorem
for finite-dimensional integrable systems. A mechanical
system with $2n$-dimensional phase space is integrable
if there are $n$ functions $F_1,\ldots, F_n$ in involution
with each other, and the Hamiltonian is a function
of $F_1,\ldots, F_n$. Then, there are $n$ action variables
$I_1,\ldots, I_n$, each of them being some combination
of $F_1,\ldots, F_n$:
$$
I_j=I_j(F_1,\ldots, F_n)
$$
such that each $I_j$ generates $U(1)$ (has periodic
orbits). 
In this paper we are dealing with an infinite-dimensional
system, a classical string in $AdS_5\times S^5$.
We can take the first Pohlmeyer charge 
$Q^{[1]}-\tilde{Q}^{[1]}$ as a Hamiltonian\footnote{It is
a natural Hamiltonian on the phase space of a classical
string in any case when the target space is a direct
product of two manifolds.}.
This Hamiltonian is presumably integrable, because
there is an infinite family of higher charges 
commuting with it. On the other hand, it does not have
any special periodicity properties (we do not see
any reason why it would). This means that the
closure of the orbit of $Q^{[1]}-\tilde{Q}^{[1]}$ is an invariant
torus. Our $U(1)_L$ commutes with $Q^{[1]}-\tilde{Q}^{[1]}$
(This fact is seen immediately, because
$Q^{[1]}$ can be rewritten as 
$\int d\tau^+\sqrt{-(\partial_+ x_A,\partial_+ x_A)}$
and by definition 
$U(1)_L$ does not act on the AdS-part of the worldsheet.)
Therefore $U(1)_L$ should be a shift of one of
the angles parametrizing the invariant torus of
$Q^{[1]}-\tilde{Q}^{[1]}$.
The angles parametrizing the invariant torus are
in correspondence with its one-dimensional cycles.
Which cycle corresponds to $U(1)_L$?
Every invariant torus can be connected by a one-parameter
family of invariant tori to a torus on the boundary
of the phase space (or the one very close to the boundary).
This means that every 1-cycle is connected to some 1-cycle
on a torus on the boundary --- the space of null-surfaces.
We should take that 1-cycle which is connected to the
orbit of $U(1)_L$ on the null-surfaces, described
in Section \ref{ChargeDefinition}.
The corresponding action variable is ${\cal E}$
--- the generator of $U(1)_L$.
These arguments show the uniqueness of $U(1)_L$.

The first Pohlmeyer charge $Q^{[1]}-\tilde{Q}^{[1]}$
has a special property: it actually generates $U(1)_L$
on the boundary. Therefore the difference between
$Q^{[1]}-\tilde{Q}^{[1]}$ and ${\cal E}$
should be a combination of charges vanishing at the boundary.
We expect
that this is an infinite and {\em linear} combination. 
Indeed, the construction of 
\cite{KT} tells us that the charge we are looking for
is local at each order in $1/|p_S|$. A nonlinear
combination of the charges would be non-local (a product
of integrals).

\subsection{A different point of view on the perturbation theory;
higher orders.}
\label{sec:Alternative}
In Section 3 we constructed $U(1)_L$ as 
$F^{-1} \circ{\partial\over\partial\psi}\circ F$
where $\partial\over\partial\psi$ is generated
by $\int d\sigma |p_S(\sigma)|$ and $F$ is the
canonical transformation such 
that $F^{-1} \circ {\partial\over\partial\psi}\circ F$
commutes with $|p_S(\sigma)|^2+|\partial_{\sigma}x_S(\sigma)|^2$. 
This canonical transformation is constructed
in the perturbation theory, order by order in $1\over |p_S|^2$.

A disadvantage of this procedure is that at each order we have
to require that our $U(1)_L$ commutes with 
$|p_S(\sigma)|^2+|\partial_{\sigma}x_S(\sigma)|^2$ {\em for any} $\sigma$.
Since there are infinitely many values of $\sigma$ we have to impose
infinitely many conditions on $F$ at each order. At the first order,
we have seen in Section \ref{StringOnASphere} that these conditions
are not really independent; one generating function $h^{(1)}$ takes
care of all of them --- see Eq. (\ref{ForAnySigma}). 
At the higher orders, this is not immediately obvious.
Therefore, we would like to propose a slightly different way of
constructing $F$. Let us forget for a moment about the
Virasoro constraint; instead of the phase space of the string
consider the space of harmonic maps $x(\tau,\sigma)$. Instead of requiring that
$U(1)_L$ commutes with $|p_S(\sigma)|^2+|\partial_{\sigma}x_S(\sigma)|^2$,
let us require that $U(1)_L$ commutes with 
$Q^{[1]}=\int d\sigma |\partial_+ x_S(\sigma)|$.
We will see that the requirement that $U(1)_L$ commutes with $Q^{[1]}$
already fixes $U(1)_L$ in the perturbation theory, and the resulting
$U(1)_L$ will automatically commute with the Virasoro constraints. 

As in Section 3, we look for the generator of $U(1)_L$ as a pullback by
a canonical transformation of $\int |p_S(\sigma)| d\sigma$.
In other words, let us look for such a canonical transformation $F$
that $\int d\sigma |p_S(\sigma)|$ commutes with $F^*Q^{[1]}$ 
(the pullback of $Q^{[1]}$ by $F$).
We can construct such a canonical transformation order by order
in the perturbation theory. 
Let us denote $K=\int d\sigma |p_S(\sigma)|$. We have:
\begin{equation}
	Q^{[1]}=K+q_1+q_2+\ldots
\end{equation}
Under the rescaling $p_S\to tp_S$: $K\to t K$, $q_1\to t^{-1} q_1$,
$q_2\to t^{-3}q_2$, $q_m\to t^{1-2m}q_m$. The symplectic structure
is of the degree 1:
$\omega\to t\omega$, therefore the Poisson brackets are of the
degree $-1$: $\{,\}\to t^{-1}\{,\}$.
We can construct $F$ order by order in this grading.
We have:
\begin{equation}
	F^*(Q^{[1]})=K+q'_1+q'_2+\ldots+q'_m+\ldots
\end{equation}
Suppose that we have already found $F$
such that $q'_1,\ldots,q'_{m-1}$ commute with $K$.
At the order $m$, we want to modify $F$ by the canonical transformation
with the generating function $f_m$ of the order $|p_S|^{1-2m}$ so that 
$q'_m+\{f_m, K\}$ commutes with $K$.
Since $K$ is periodic, we can decompose 
\begin{equation}
	q'_m=q'_{m,0}+\sum_{k\neq 0} q'_{m,k}
\end{equation}
where $\{K,q'_{m,k}\}=ik q'_{m,k}$.
Then we should take
\begin{equation}
	f_m=\sum_{k\neq 0} {1\over ik}q'_{m,k}
\end{equation}
Repeating this procedure at higher orders, we end up with the
function $F$ such that $\{K, F^*(Q^{[1]})\}=0$.

The reparametrization invariance
is manifestly preserved at each order, therefore the resulting
charge $F^{-1} \circ{\partial\over\partial\psi}\circ F$
will commute with $(p_S,\partial_{\sigma}x_S)(\sigma)$
for any $\sigma$.
Also, the fact that $Q^{[1]}$ is reparametrization-invariant and 
the arguments analogous to the discussion 
at the end of Section \ref{sec:Commutes}
show that $F^{-1} \circ{\partial\over\partial\psi}\circ F$ 
will automatically
commute with $|p_S(\sigma)|^2+|\partial_{\sigma}x_S(\sigma)|^2$,
as well as with the higher Pohlmeyer charges.
Indeed, we know that $F^*Q^{[1]}=K+q_1'+q_2'+\ldots$ commutes
with $F^*(|p_S|^2(\sigma)+|\partial_{\sigma}x_S|^2(\sigma))=
|p_S|^2+\phi_0+\phi_1+\ldots$;
therefore 
\begin{eqnarray}
	&& \{K, \phi_0\}=\{|p_S|^2(\sigma),q_1'\} \Rightarrow 
	\{K, \{K, \phi_0\}\}=0\Rightarrow \{K,\phi_0\}=0\nonumber\\[5pt]
	&& \{K, \phi_1\} + \{q_1',\phi_0\}=
	\{|p_S|^2(\sigma),q_2'\}\Rightarrow \{K, \{K, \phi_1\}\}=0
	\Rightarrow \{K, \phi_1\}=0\nonumber\\[5pt]
	&& etc.\nonumber
\end{eqnarray}
We used the periodicity of the trajectories of $K$ when
we claimed that $\{K, \{K, \phi\}\}=0$ implies
$\{K,\phi\}=0$. Indeed, for any functional $\phi$ on the
phase space, if $\{K, \{K, \phi\}\}=0$ then 
$\{K,\phi\}$ is constant on the trajectories of $K$.
But if this constant were nonzero, then the change 
of $\phi$ along the trajectory of $K$ would accumulate
over the period  of $K$, which would contradict
the single-valuedness of $\phi$ on the phase space.

\subsection{An infinite combination of local conserved charges.}
Expanding (\ref{ChargeUV}) in the conformal gauge in the 
powers of $1\over |p_S|$ we get:
\begin{eqnarray}
	&&{1\over 2}\left(Q^{[2]}-\widetilde{Q}^{[2]}\right)
	=
	\int d\sigma \left[
	-2|p_S|+{3\over |p_S|}(\partial_{\sigma}x_S)^2
	-\right.\\[5pt]&&\left.
	-{3\over |p_S|^3}(p_S,\partial_{\sigma}x_S)^2
	+{4\over |p_S|}\left(D_{\sigma}{p_S\over |p_S|},
	D_{\sigma}{p_S\over |p_S|}\right)+\ldots\right]
	\nonumber
\end{eqnarray}
And for $Q^{[1]}$ we get:
\begin{equation}\label{Ecal2}
	Q^{[1]}-\widetilde{Q}^{[1]}=
	\int d\sigma \left[
	2|p_S|+{1\over |p_S|}(\partial_{\sigma}x_S)^2
	-{1\over |p_S|^3}(p_S,\partial_{\sigma}x_S)^2
	+\ldots\right]
\end{equation}
We have:
\begin{eqnarray}
&&	{1\over 16}\left[7(Q^{[1]}-\widetilde{Q}^{[1]})-
{1\over 2}(Q^{[2]}-\widetilde{Q}^{[2]})\right]=
	\nonumber\\[5pt]=
	&&	\int d\sigma \left[|p_S|+{1\over 4|p_S|}
	\left[ (\partial_{\sigma}x_S)^2-
\left(D_{\sigma}{p_S\over |p_S|},
	D_{\sigma}{p_S\over |p_S|}\right)
	-{(p_S,\partial_{\sigma}x_S)^2\over (p_S,p_S)}
	\right]+\ldots
	\right]
	\nonumber
\end{eqnarray}
This coincides with the result (\ref{PeriodicSecondOrder})
for ${\cal E}$ which we know from the perturbation theory.
We see that up to the terms of the order $1\over |p_S|^3$
the Hamiltonian of $U(1)_L$ can be represented as 
a sum of the first two commuting local charges.
We conjecture that $U(1)_L$ is in fact
an infinite combination of the local conserved charges.
The perturbation theory construction suggests
that it should be a worldsheet parity-invariant combination.

The coefficients of this linear combination can be 
found from considering the conserved charges of 
particular solutions. There is a special class of 
fast moving strings, the so-called ``rigid'' strings.
For these ``rigid'' strings, the corresponding 
field theory operators are known a~priori. These
operators provide local extrema of the anomalous dimension
in the sector with the given charges.
These ``rigid'' solutions were classified
in \cite{SpeedingStrings,ART}. They are related
to the solutions of the Neumann integrable system.
The local conserved charges of some rigid strings 
were computed in \cite{ArutyunovStaudacher,Engquist}.

In \cite{ArutyunovStaudacher} the local conserved charges are
denoted ${\cal E}_k$. (This agrees with our notation
${\cal E}$ for the Hamiltonian of $U(1)_L$.)
The precise definition of ${\cal E}_k$ is given in Section 3
 of \cite{ArutyunovStaudacher}. The relation to our
 notations is: $Q^{[1]}-\tilde{Q}^{[1]}=2{\cal E}_2$, 
 $Q^{[2]}+2Q^{[1]}-\tilde{Q}^{[2]}-2\tilde{Q}^{[1]}=-4{\cal E}_4$. 
The conserved charges have the following 
structure:
\begin{equation}
	{\cal E}_n=\delta_{2,n}{\cal J} 
	+{\epsilon^{(1)}_n\over {\cal J} }
	+{\epsilon^{(2)}_n\over {\cal J}^3 }
	+{\epsilon^{(3)}_n\over {\cal J}^5}+\ldots
\end{equation}
where  ${\cal J}^{-2}=\lambda/ J^2$, and $J$ is a particular
combination of the $SO(6)$ momenta. 
The coefficients $\epsilon_n^{(m)}$ depend on what kind of
a rigid string is considered (the ratio of spins).
But the authors of \cite{ArutyunovStaudacher} noticed
that the coefficients $\epsilon_n^{(m)}$ for different
values of $n$ are not independent. For all the solutions they
considered, they find that:
\begin{equation}
	{\cal E}_{10}+{74\over 7}{\cal E}_8+
	{1898\over 35}{\cal E}_6+
	{6922\over 35}{\cal E}_4+
	{32768\over 35}({\cal E}_2-{\cal J})\sim
	{1\over {\cal J}^9 }
\end{equation}
This means that up to the terms of the order $1/|p_S|^9$
we should have:
\begin{equation}\label{JFromCalE}
	{\cal J}=
	{\cal E}_2+{6922\over 32768}{\cal E}_4+
	{1898\over 32768}{\cal E}_6+
	{370\over 32768}{\cal E}_8+
	{35\over 32768}{\cal E}_{10}+\ldots
\end{equation}
At first this formula looks rather strange, because it
seems to imply that a certain combination of  Pohlmeyer
charges (which all commute with $SO(6)$) is equal to 
some component of the angular momentum (which transforms in the adjoint
of $SO(6)$).  We propose the following resolution
of this puzzle. The right hand side of (\ref{JFromCalE})
is actually the action variable, which for 
a particular class of the solutions considered in
\cite{ArutyunovStaudacher,Engquist} happens to be equal to 
the $SO(6)$ charge $J$ (because these particular
solutions correspond to the chiral
operators on the field theory side; see Section 2 of
\cite{SlowEvolution}).
In other words, for this particular class of solutions
the angular momentum 
${\cal J}$ should be equal to our action variable ${\cal E}$. 
The general formula should have on the left hand side
${\cal E}$, the generator of $U(1)_L$, instead of ${\cal J}$:
\begin{equation}\label{LHSisAction}
	{\cal E}=
	{\cal E}_2+{6922\over 32768}{\cal E}_4+
	{1898\over 32768}{\cal E}_6+
	{370\over 32768}{\cal E}_8+
	{35\over 32768}{\cal E}_{10}+\ldots
\end{equation}
This gives the expansion of the generating function
of $U(1)_L$ to the order $1\over |p_S|^9$.
It would be interesting to check explicitly, beyond the
order $1/|p_S|$, that this Hamiltonian generates periodic
trajectories.

\subsection{More on the perturbation theory.}
Here we want to present a slightly different and perhaps simpler
way of thinking
about the continuation of $U(1)_L$ in the perturbation theory.
Consider the Hamiltonian vector field $\xi_{ {\cal E}_2}$
corresponding to the first Pohlmeyer charge ${\cal E}_2$.
Consider the canonical transformation
\begin{equation}
	F=e^{2\pi \xi_{ {\cal E}_2}}
\end{equation}
This canonical transformation is the Hamiltonian flow
generated by ${\cal E}_2$ by the time $2\pi$. 
The trajectories of ${\cal E}_2$ are almost periodic
in the null-surface limit, therefore we can write
\begin{equation}
	F=e^{v_1}
\end{equation}
where $v_1$ is a vector field of the order
$1/|p_S|^2$. This vector field can be constructed in the following way.
Let us choose the conformal gauge on the worldsheet.
We know from (\ref{Ecal2}) that 
${\cal E}_2=\int d\sigma |p_S|+ f=K+f$
where $f$ is of the order $1/|p_S|$.
Taking into account that $e^{2\pi \xi_K}={\bf 1}$ we get
\begin{eqnarray}
&&	F={\bf 1}+\int_0^{2\pi} ds 
	e^{- s\xi_K}\xi_{f}e^{s\xi_K}+\nonumber \\
&&	+\int_{s_1<s_2} ds_2 ds_1\;
	e^{- s_2\xi_K}\xi_{f} e^{s_2\xi_K}
	e^{-s_1\xi_K}\xi_{f} e^{s_1\xi_K}+\ldots= \nonumber \\
	&&	=\exp\left\{ \int_0^{2\pi} ds e^{-s\xi_K}\xi_{f}e^{s\xi_K}
	+\right.\\ &&\left.+{1\over 2}\int_{s_1<s_2}ds_1 ds_2 \;
	[e^{- s_2\xi_K}\xi_{f} e^{s_2\xi_K},
	e^{-s_1\xi_K}\xi_{f} e^{s_1\xi_K}]+\ldots
	\right\}
	\nonumber
\end{eqnarray}
This defines 
$$v_1=\int_0^{2\pi} ds e^{-s\xi_K}\xi_{f}e^{s\xi_K}
	+{1\over 2}\int_{s_1<s_2}ds_1 ds_2 \;
	[e^{- s_2\xi_K}\xi_{f} e^{s_2\xi_K},
	e^{-s_1\xi_K}\xi_{f} e^{s_1\xi_K}]+\ldots$$
in the perturbation theory.
(Notice that $f$ can be decomposed in the Fourier series
$f=\sum f_k$ so that $\{K,f_k\}=ikf_k$ and then the leading
term of $v$ is the zero mode $\xi_{f_0}$; this is the
``averaging'' procedure of \cite{SlowEvolution}.)
The vector field $v_1$ defines the vector field on
the moduli space of null-surfaces as a limit $\lim_{|p_S|\to\infty}
({\cal E}_2^2 v_1)=\lim_{|p_S|\to\infty}(L^2v_1)$ (where $L$ was
defined in Section \ref{sec:Parametrization}). 
This vector field determines the
slow evolution of the null-surface;
it is the Hamiltonian vector field of the
Landau-Lifshitz model on the moduli 
space of the null-surfaces modulo $U(1)_L$ \cite{SlowEvolution,SNS}.

As in \cite{ArutyunovStaudacher} we can consider the
improved currents ${\cal E}'_{2n}$. By definition
${\cal E}'_{2n}$ is a linear combination of 
${\cal E}_2,\ldots,{\cal E}_{2n}$ such that 
${\cal E}'_{2n}=O(|p_S|^{-2n+3})$.
The Hamiltonian of the Landau-Lifshitz model 
is the null-surface limit of ${\cal E}'_4$,
more precisely $\lim_{|p_S|\to\infty}({\cal E}_2 {\cal E}'_4)$.
Given that ${\cal E}_2$ and ${\cal E}'_4$ are in involution,
this implies that for some $a_1$ we have  
\begin{equation}
	F_1=e^{2\pi(\xi_{ {\cal E}_2}+a_1\xi_{ {\cal E}'_4})}=e^{v_2}
\end{equation}
where $v_2$ is of the order $1/|p_S|^4$.
Again, $\lim_{|p_S|\to\infty}({\cal E}_2^4 v_2)$ determines
a  vector field on the moduli space of null-surfaces. (It can be
also defined as $\lim_{|p_S|\to\infty}(L^4 v_2)$.)
This vector field commutes with the time evolution of the
Landau-Lifshitz model.
We conjecture
that this vector field is generated by the second conservation
law of the Landau-Lifshitz model, which is proportional
to the null-surface limit\footnote{Notice that
the null-surface limit of ${\cal E}_{2n}'$ is
invariant under the $U(1)_L$ symmetry of the
null-surfaces, because the
$U(1)_L$ symmetry of the null-surfaces is generated
by  the conserved quantity ${\cal E}_2$ 
which commutes with ${\cal E}'_{2n}$.}
of ${\cal E}_6'$, more precisely
 $\lim_{|p_S|\to\infty}({\cal E}_2^3 {\cal E}'_6)$
or $\lim_{|p_S|\to\infty}(L^3{\cal E}'_6)$.
Repeating this procedure we
get 
$$
e^{2\pi(\xi_{ {\cal E}_2}+a_1\xi_{ {\cal E}'_4}+
a_2\xi_{ {\cal E}'_6}+\ldots)}={\bf 1}
$$
in the perturbation theory. These arguments lead us to the 
following conclusion.
First, we see once again that there is a linear 
combination $ {\cal E}_2+a_1 {\cal E}'_4 +
a_2{\cal E}'_6+\ldots $ generating periodic trajectories.
Second, the moduli space of null-surfaces
in $AdS_5\times S^5$ modulo $U(1)_L$ is naturally equipped
with the infinite tower of Hamiltonians
in involution which are the null-surface limit of the
Pohlmeyer charges. This is the generalized Landau-Lifshitz model.
	
\section{Conclusion.}
Given a manifold with the metric of the Lorentzian signature
it is possible to construct the extremal surfaces in this manifold 
as perturbations
of the null-surfaces. In the special case when the manifold
is $AdS_5\times S^5$ the AdS/CFT correspondence predicts that
the extremal surfaces (which are the same as classical string
worldsheets) correspond to the states of the large
R-charge in the ${\cal N}=4$ supersymmetric Yang-Mills theory.
From this point of view considering the extremal surface
as a perturbation around the null-surface corresponds to 
considering the state of the interacting Yang-Mills theory as 
a perturbation of the state of the free Yang-Mills theory.
This correspondence has the following important features:
\begin{enumerate}
	\item Locality.
In the planar limit (the limit of infinitely many colors) the
Yang-Mills perturbation theory is local in the following sense:
the Feynman diagramms involve only intractions of
those elementary field operators which stand next to each other in the
product under the trace. 
We expect that the correspondence between the parton chains
and the string worldsheets is local in each order of the
perturbation theory, and therefore the locality of the planar
Yang-Mills perturbation theory should correspond to the
locality of the string worldsheet theory. 
        \item Integrability.
The classical string worldsheet theory in $AdS_5\times S^5$ 
is an integrable system.
\end{enumerate}
Because of the integrability, there is an infinite family
of local conserved charges in involution. In this paper we have argued
that an infinite linear combination of these local charges
generates periodic trajectories on the string phase space. 
This statement can be verified order by order in the null-surface
perturbation theory, and it is local at each order.
This means that the ``slow evolution'' of nearly-degenerate
extremal surfaces \cite{SlowEvolution} is essentially controlled
by the Pohlmeyer charges (we will further discuss 
the slow evolution and how it is related to the Pohlmeyer charges
in the second paper of \cite{Later}). 

It would be interesting to further study the null-surface perturbation theory
from the point of view of the integrability. It would be especially
interesting 
to study those manifestations of the integrability
which are local. 
The B\"acklund transformations \cite{Pohlmeyer} is one example. They allow us
to construct a new extremal surface from a given extremal surface,
and in the null-surface perturbation theory these transformations are 
well-defined and local at each order.
The B\"acklund tranformations
are closely related to the local conserved charges, 
and in fact the hidden symmetry $U(1)_L$
can be considered as a consequence of the special properties
of these transformations. We will discuss the relation 
between $U(1)_L$ and
the B\"acklund transformations in the third paper of \cite{Later}.

The general problem is, to study those aspects of the integrability
which are local 
in the null-surface perturbation theory.
(Without a reference to the null-surface perturbation theory,
we would define the locality as some sort of an independence 
of the choice of the boundary conditions.)
This problem arises also on the field theory side. 
The Feynman diagramms in the planar limit are local, but we usually compute
the anomalous dimension of the single-trace operators which requires 
summing over the whole parton chain. The spectrum of
single-trace operators at large $N$ is certainly an invariant of the theory,
but it is non-local. 
If it is true that the planar ${\cal N}=4$ Yang-Mills theory is integrable,
it would be important to understand the integrability as much as possible
in terms of the local properties of the parton chain (perhaps
on the level of the individual Feynman diagramms).

We have defined the $U(1)_L$ strictly speaking in the perturbation theory, 
but it should be actually well-defined in the domain of the string phase
space where the velocity of the string is large enough.
In other words, the series defining the $U(1)_L$ in fact converges
if the string moves fast enough.  
It would be interesting to study the global
properties of $U(1)_L$.

An important question is what happens to $U(1)_L$ after the quantization.
To answer this question we should first 
include fermions. Important steps in this direction
were made recently in 
\cite{ArutyunovFrolov0411,BKSZ,AAT}.

It would be interesting to understand better why the ``length''
is conserved on the field theory side. (Why
is there a quantum number $L$ with a well-defined classical limit?)
To which extent the conservation of $L$ is related to the integrability
of the planar Yang-Mills theory? 
What happens to $L$ when we turn on the fermions?

Null-surfaces are obviously an important ingredient in our construction.
The correspondence between the null-surfaces and the ``engineering''
operators in the free field theory is rather straightforward; 
the null-surfaces in $AdS_5\times S^5$ appear
very naturally in the description of the coherent states of the
free theory \cite{Kruczenski,SNS}. 
Is there any way to see directly on the field theory side, that
turning on the Yang-Mills interaction  corresponds to the
deformation of the null-surface into the extremal surface?

\section*{Acknowledgments}
I would like to thank  S.~Frolov, A.~Gorsky, V.~Kaloshin, A.~Kapustin,
A.~Marshakov,
S.~Minwalla,
M.~Van~Raamsdonk, A.~Starinets, K.~Zarembo and especially A.~Tseytlin
for discussions, and G.~Arutyunov
for a correspondence on the local conserved
charges. I want to thank A.~Tseytlin for comments on the text. 
I want to thank the organizers
of the String Field Theory Camp at the Banff International 
Research Station for their hospitality while this 
work was in progress. 
This research was supported by the Sherman Fairchild 
Fellowship and in part
by the RFBR Grant No.  03-02-17373 and in part by the 
Russian Grant for the support of the scientific schools
NSh-1999.2003.2.

\end{document}